\documentclass[aps,prl,twocolumn,superscriptaddress,showpacs]{revtex4}
\usepackage{graphicx}
\usepackage{xcolor}
\usepackage{amsmath,amssymb,latexsym}
\usepackage{ulem}
\usepackage{mathrsfs}
\newcommand{\RNum}[1]{\uppercase\expandafter{\romannumeral #1\relax}}

\begin{document}

\title{Calibrating the standard candles with strong lensing}

\author{Xudong Wen}
\author{Kai Liao}\email{liaokai@whut.edu.cn}
\affiliation{School of Science, Wuhan University of Technology, Wuhan 430070, China}

\begin{abstract}
We propose a new model-independent strategy to calibrate the distance relation in Type Ia supernova (SN) observations and to probe the intrinsic properties of SNe Ia, especially the absolute magnitude $M_B$, basing on strong lensing observations in the upcoming Large Synoptic Survey Telescope (LSST) era. The strongly lensed quasars can provide the Time Delay Distances (TDD) and the Angular Diameter Distances (ADD) to the lens galaxies. These absolute distance measurements can model-independently anchor the SNe Ia at cosmological distances. We simulated 55 high-quality lensing systems with $5\%$ uncertainties for both TDD and ADD measurements basing on future observation conditions. For the time delay distances and the angular diameter distances as the calibration standards, the calibrated $1\sigma$ uncertainties of $M_{B}$ are approximately 0.24 mag and 0.03 mag, respectively. Besides, we also consider an evolving distance relation, for example, caused by the cosmic opacity. In this case, the $1\sigma$ uncertainties of $M_B$ calibrated with TDD and ADD are approximately 0.31 mag and 0.06 mag, respectively. The results show that the ADD method will be a promising tool for calibrating supernovae.
\end{abstract}

\date{\today}
\maketitle

\section{Introduction}

Over the past decades, the number of observed Type Ia supernovae (SNe) have increased dramatically with the development of large surveys. Cosmography is therefore better set up.
SNe Ia are extremely luminous explosions and have almost the same peak absolute magnitude ($M_B$) on the basis of a reasonable physical mechanism~\cite{Hoyel}.
For most ``\,normal\,'' SNe Ia, their peak absolute magnitudes have small dispersion~\cite{Branch}.  They are therefore taken as ideal standard candles in determining extra-galactic and cosmological luminosity distances.
However, the value of the peak absolute magnitude (i.e., intrinsic brightness) is unknown and considered as a free parameter, which needs to be calibrated by the local distance ladders.

In practice, the way to calibrate the SNe Ia is through
Cepheid variable stars at local Universe whose
luminosities are related with periods~\cite{Branch1998}.
This distance ladder method has led to good results in cosmological studies~\cite{Riess2009,Jones}. However, there are some uncertainties in this approach. Firstly, it mainly depends on the period-luminosity relationship of Cepheid variables~\cite{Phillipse}. The effect of metallicity on both the zero-point and slope of this relationship is highly controversial in different theories. The effects of photometric contamination and a changing extinction law on Cepheid distances are also uncertain factors that cannot be ignored~\cite{Freedman}.
Secondly, with the development of the SN theory, many new factors have been found to be related to the absolute magnitude of supernovae. For example, Host galaxies in different evolutionary stages have different roles in the accretion process, resulting in the absolute luminosity of supernovae dependent on environments~\cite{Gilfanov}. For the formation of supernovae, there are many possibilities for the nature of the companion star of the white dwarf. The single-degenerate path has been successful in explaining the observations of the SNe Ia~\cite{Hachisu}. But in the case of a double-degenerate path, subluminous SNe Ia that are dimmer than their typical counterparts are produced ~\cite{Pakmor}.
Thirdly, the Cepheid variable stars are measured locally, and the calibration needs to be extrapolated to high redshifts. Considering that a high redshift supernova is redder and more massive than a low redshift supernova~\cite{Amanullah}, some relations between the properties of supernovae may change as the redshift increases. Therefore, it cannot be checked whether this extrapolation method is effective in the case of high redshifts. Due to the potential absorption, scattering of the photons or other mechanisms that
transfer photons to other particles~\cite{Liao_tau}, the cosmic opacity could make the SNe Ia dimmer, equivalently making $M_B$ appear to evolve with redshift.

Recently, the community is puzzled by the Hubble constant ($H_0$) tension issue. The $H_0$ measured from Cepheid variable stars and SNe Ia at local Universe
has $4.4\sigma$ mismatch with that constrained from Cosmic Microwave Background (CMB) observations in the standard $\Lambda$CDM model~\cite{Riess2019}.
This contradiction would either manifest unknown systematic errors in local distance measurements and CMB observations measurements or imply new physics beyond the standard cosmological model.
Note that besides the Cepheid calibration, one can use a cosmological model to calibrate
the SNe Ia at high redshifts by simultaneously fitting the parameters in the model and
parameters of SNe Ia, for example, $M_B$~\cite{Betoule}.

Therefore, due to the issues about both systematic errors in observations and the standard cosmological model mentioned above, it is necessary to explore new model-independent calibration methods (even at cosmological distances to achieve the cross-check). It should be emphasized here that there are at least three benefits to do this: 1) understand the properties of SNe Ia themselves at any redshifts directly and cosmological-model-independently; 2) provide new ways to anchor SNe Ia and then apply them in cosmological studies; 3) the newly calibrated SNe Ia may shed light on the $H_0$ tension issue.

In the literature, the effective absolute magnitude $M(z)$ was calibrated by using the Etherington's distance-duality relation and the angular baryonic oscillation (BAO) scale observed at any redshifts~\cite{Evslin}. The disadvantage of this method is that it produces the quantity that cannot be directly compared with SN simulations.
The inverse distance ladder technique can not only determine $H_0$ but also calibrate supernovae.
Some articles use this method to calibrate the intrinsic magnitude of supernovae by combining supernovae and BAO~\cite{Macaulay,Aubourg2015}.
However, BAO relies on the scale of the sound horizon at recombination $r_s$ to convert angular measurements into angular-diameter distance~\cite{Wojtak}. This means that once the $r_s$ is fixed the $H_0$ has already been determined.
Recently, the value of $H_0$ was determined by using the inverse distance ladder method in combination with supernovae and gravitational lenses~\cite{Taubenberger}, though it depends on a specified cosmological model. Another study used three time-delay lenses to calibrate the distance ladder at low-redshifts, combined them with relative distances from SNe Ia and BAO, leaving $r_s$ completely free~\cite{Wojtak}. This method calibrates supernovae from a new perspective and is promising in the future.
Moreover, the discovery of a coalescing gravitational wave (GW) signal of a compact binary system and its electromagnetic counterpart provides a new method for calibrating supernova absolute magnitudes~\cite{zhao,Gupta}. It is expected that the third generation of gravitational wave detectors will provide more abundant data in the future.

Strong gravitational lensing has become an effective tool in astrophysics and cosmology~\cite{Treu}. When light from a distant object passes through an elliptical lens galaxy, multiple images of AGN can be observed and time delays exist among them due to the geometric and Shapiro effects for different paths. Distances can be obtained by analyzing the imaging and time delays. There are two methods to extract distance information. One is to measure the ``\,time delay distance\,'' (TDD) consisting of three angular diameters distance~\cite{liao}. The other is to measure the angular diameter distance (ADD) of the lenses, which can be obtained by measuring the time delays and the velocity dispersion of the lens galaxy~\cite{Paraficz,Jee}.
The current and upcoming large surveys are bringing us a large number of lensed quasars, making time delay measurements of strong lensing systems very promising for cosmology.

In this work, we propose to use two kinds
of lensing distances for calibrating the SNe Ia at cosmological distances. It should be noted that the lensing observations are angular separation and spectroscopy measurements, thus the distances measured should be free of cosmic opacity~\cite{liao2019}.
The structure of the paper is as follows. In Sect.~\ref{Sec2} we introduce the angular diameter distance and time delay distance, respectively. We also introduce the mock catalog of the strong lensing systems. In Sect.~\ref{Sec3} We introduce the method for calibrating SNe Ia with or without the consideration of cosmic opacity. In Sect.~\ref{Sec4} we present our analysis and results. Finally, we summarize our work in Sect.~\ref{Sec5}.

\section{Distances from Strong Lensing}
\label{Sec2}

Thousands of lensed quasars will be detected by the upcoming wide-field synoptic surveys. In particular, the Large Synoptic Survey Telescope (LSST)~\cite{Ivezic} will find more than 8000 lensed quasars, of which a considerable part have well-measured time delays~\cite{Oguri}. With ancillary data consisting of high-quality imaging from
next generation space telescope, the central velocity dispersion of the lens galaxies
and the line-of-sight (LOS) measurements, we can measure the
TDD and ADD. We introduce both of them in the following.
To make it clearer, we take the Singular Isothermal Sphere (SIS)~\cite{Keeton} as the model of the lens for example,
although realistic lenses are much more complicated.

Firstly,
The time delay between two images of the lensed AGN can be expressed by an equation containing TDD as:
\begin{equation}
    \Delta t=\frac{D_{\Delta t}}{c}\Delta\phi,
\end{equation}
where c is the light speed. $\Delta\phi$ is the difference between Fermat potentials at different image positions, which can be inferred by high resolution imaging observations of Einstein ring (or arcs). In the SIS model, $\Delta \phi=(\theta_i^2-\theta_j^2)/2$~\cite{Jee}.
The TDD is defined by:
\begin{equation}
  D_{\Delta t}=(1+z_{l})\frac{D^{A}_{l}D^{A}_{s}}{D^{A}_{ls}},
\end{equation}
which is the combination of three different angular diameter distances~\cite{Suyu2010}. $D^{A}_{l}, D^{A}_{s}, D^{A}_{ls}$ are the angular diameter distances between observer and lens, observer and source, and lens and source, respectively. $z_l$ is the lens redshift. Therefore, if we obtain the time delay through monitoring the light curves and model the potential of the lens, we can get the TDD.

Secondly,the random motion of stellar in an elliptical galaxy produces Doppler shift on the spectra corresponding to each stellar, and the velocity dispersion $\sigma$ can be obtained by observing the integrated spectrum of the whole galaxy. From the Virial theorem, $\sigma$ is related to the mass $M_{\sigma}$ in radius R, $\sigma^{2}\propto M_{\sigma}/R$~\cite{Paraficz}. In a gravitational lens system, the relationship between Einstein angle $\theta_{E}$ and mass $M_{\theta_{E}}$ is as follows:
\begin{equation}
  \theta_{E}=\sqrt{\frac{4GM_{\theta_{E}}}{c^{2}}\frac{D^{A}_{ls}}{D^{A}_{l}D^{A}_{s}}},
\end{equation}
where the radius of the Einstein ring can be expressed as $R= D^{A}_{l}\theta_{E}$. Therefore, it can be deduced that: $\sigma^{2}\propto\frac{D^{A}_{s}}{D^{A}_{ls}}\theta_{E}$. In the SIS model, velocity dispersion is given by~\cite{Jee}
\begin{equation}
  \sigma^{2}=\theta_{E}\frac{c^{2}}{4\pi}\frac{D^{A}_{s}}{D^{A}_{ls}}.
\end{equation}
Considering that $\Delta t$ is proportional to $\frac{D^{A}_{s}D^{A}_{l}}{D^{A}_{ls}}$ and the velocity dispersion $\sigma^2$ is proportional to $\frac{D^{A}_{s}}{D^{A}_{ls}}$, the angular diameter distance $D^{A}_{l}$ to the lens can be obtained by the ratio $\Delta t/\sigma^{2}$~\cite{Paraficz}. In a SIS lens, angular diameter distance $D^{A}_{l}$ can be written as
\begin{equation}
  D^{A}_{l}=\frac{c^{3}}{4\pi}\frac{\Delta t}{\sigma^{2}(1+z_{l})}.
\end{equation}

\begin{figure}
\includegraphics[width=8cm,angle=0]{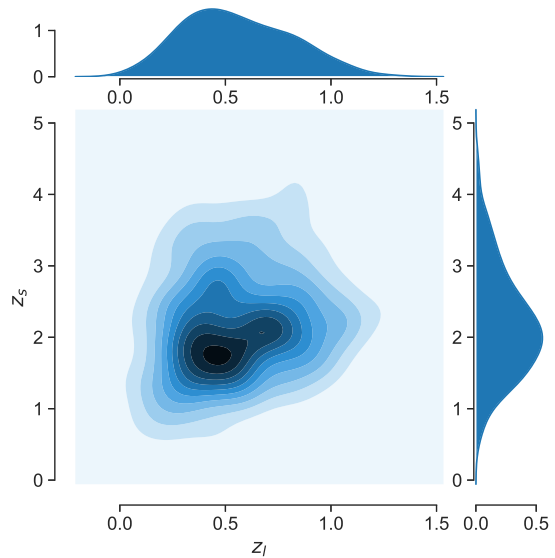}
\caption{The lens and source redshift distributions of the lens systems with well-measured time delay light curves observed by LSST plus excellent auxiliary data such that the measured distances have 5\% precision. For visualization, we show it in the form of probability density}
\label{result}
\end{figure}

The Time Delay Challenge (TDC) program tested the accuracy of current algorithms~\cite{Liao2015}. And with the first challenge (TDC1), the average precision of the time delay measurement was approximately $\sim3\%$, which was comparable to the uncertainty of current lens modeling~\cite{Suyu}. Considering that the metric efficiency was about 20\%, TDC1 gave at least 400 well-measured time delay systems~\cite{Liao2015}. Since the TDD and ADD are sensitive to the mass distribution of the lens, auxiliary data such as high-resolution imaging and stellar velocity dispersion observations are required for accurate lens modeling, so that reasonably accurate distance information can be obtained.
By setting the selection criteria:  (1) the angular separation of lensed images is $>1''$~\cite{Jee2016}, (2) the third brightest quasar image has an i-band magnitude $m_{i}<21$~\cite{Jee2016}, (3) the lens galaxy has $m_{i}<22$~\cite{Jee2016}, (4) considering the quadruple imaging lens systems that provide more information than the double imaging systems, for example, the general Source-Position Transformation (SPT) does not conserve the time delay ratios in some cases
\cite{Schneider,Schneider2} (note the Mass-Sheet Transformation conserves the ratios). One can see this clearly in the H0LiCOW samples~\cite{Wong}. SDSS 1206+4332 is a double-imaged
system which yields weaker constraint on $H_0$ even though the host galaxy is quadruply-imaged providing
additional constraints for the lens modeling.
In the end, we will have $\sim$55 high-quality quad-image the angular separation of lensed images in the mock catalog\cite{Jee2016}.
As in Jee et al. 2016, we set $5\%$ uncertainties for both TDD and ADD (see also ~\cite{liao,Linder2011}).
We plot the redshift distributions of the lenses and sources that match the selection criteria in Fig.\ref{result}, and randomly generate 55 samples from it.

\section{Methodology}
\label{Sec3}

For SN Ia data, we use a catalog of direct SN Ia observations: a joint analysis of SN Ia observations obtained by the SDSS-II and SNLS collaborations. The catalog includes several low-redshift samples, three seasons from SDSS-\RNum{2} $0.05<z<0.4$, and three years of data from SNLS $(0.2<z<1)$. It contains in total 740 spectroscopically confirmed type Ia supernovae with high quality light curves. This data set is called ``\,joint light curve analysis\,'' (hereinafter referred to as JLA)~\cite{Betoule}.

A modified version of the Tripp formula can
transform SALT2 light-curve fit parameters to distance
modulus~\cite{Tripp}:
\begin{equation}
\mu(\alpha,\beta,M_B)=m_B-M_B+\alpha x-\beta c, \label{mub}
\end{equation}
where $m_B$ is the rest-frame peak magnitude in the B band, $x$ is the stretch determined by
the shape of the SN Ia light curve and $c$ is the color measurement. $\alpha$ and $\beta$ are nuisance parameters that characterize stretch-luminosity and color-luminosity relationships. $M_B$ is also a nuisance parameter standing for the B band absolute magnitude. Further, we use a procedure mentioned in Conley et al.2011 (C11)~\cite{Conley2011} that can approximately correct the effect of the host stellar mass $(M_{stellar})$ on the intrinsic luminosity of the SNe Ia by a simple step function~\cite{Betoule}:
\begin{equation}
M_{B} = \begin{cases}
M_{B}^{1}, &\text{if }  M_{stellar }<10^{10}M_{\odot}.\\
M_{B}^{1}+\Delta_{M}, &\text{otherwise}.

\end{cases}
\end{equation}
For robustness and simplicity, we only consider the statistical uncertainties.
The error of the distance modulus $\mu$ can be expressed as:
\begin{equation}
    \sigma_{\mu}=\sqrt{\sigma_{m_{B}}^{2}+\alpha^{2}\sigma_{x}^{2}+\beta^{2}\sigma_{c}^{2}},
\end{equation}
where $\sigma_{m_{B}}$, $\sigma_{x}$, and $\sigma_{c}$ are the errors of the peak magnitude $m_B$ and light curve parameters $(x,c)$ of the SNe Ia, respectively.

The luminosity distance of SN Ia in Mpc can be obtained by
\begin{equation}
 D_{SN}^{L}=10^{\mu/5-5}.
\end{equation}
To compare distances of SNe Ia with lensing distances, we need to use the corresponding
angular diameter distances $D^{A}_{SN}$
from SNe Ia which can be easily obtained through the Distance Duality Relation (DDR)~\cite{Etherington}:
\begin{equation}
  D^{A}_{SN}=\frac{D^{L}_{SN}}{(1+z)^{2}},
\end{equation}
where the error of $D^{A}_{SN}$ can be expressed as
\begin{equation}
    \sigma_{D^{A}_{SN}}=(\ln{10}/5)D_{SN}^{A}\sigma_{\mu}.
\end{equation}

By using Eq. 10, the angular diameter distances from the observer to the lens ($D^{A}_{SN,l}$) and from the observer to the source ($D^{A}_{SN,s}$) can be obtained, respectively. Considering that in the flat universe case, the comoving distance $r=(1+z)D^{A}$ between the lens and the source can be written as $r_{ls}=r_{s}-r_{l}$~\cite{Liao2016}. Thus, the angular distance from the lens to the source $D^{A}_{SN,ls}$ is given by:
\begin{equation}
D^{A}_{SN,ls}=D^{A}_{SN,s}-\frac{1+z_{l}}{1+z_{s}}D^{A}_{SN,l}.
\end{equation}
Then we can construct the TDD from SN Ia observations
\begin{equation}
D_{\Delta t, SN}=\frac{D^{A}_{SN,l}D^{A}_{SN,s}}{D^{A}_{SN,ls}},
\end{equation}
with the corresponding error being
\begin{equation}
\sigma_{D_{\Delta t, SN}} = \sqrt{\sigma_{{D^{A}_{SN,s}}}^{2}\left(\frac{\partial D_{\Delta t, SN}}{\partial D^{A}_{SN,s}}\right)^{2}+\sigma_{{D^{A}_{SN,l}}}^{2}\left(\frac{\partial D_{\Delta t, SN}}{\partial D^{A}_{SN,l}}\right)^{2} }.
\end{equation}

In order to perform the calibration, in principle, the distances from two kinds of data should correspond to the same redshift. However, their redshift cannot always be matched perfectly. One solution is to select nearby data pair whose redshift difference is small enough to be considered the same. In this paper, we use redshift difference $\Delta z<0.005$ as the screening criterion~\cite{liao2019}. If there are more than one point in the range of screening criterion, the one with the smallest $\Delta z$ is chosen.

We consider that this work does not attempt to give an accurate constraint on parameters of SNe from the realistic data, but to propose a new method and give an estimate of the precision level based on data simulated by future observation conditions.
In the actual data, the bias is assessed by the non-Gaussian effect, which has to start from the original observations, that is, the pixel values of the host imaging, the velocity dispersion, the AGN positions, and the time delays taken as Gaussians~\cite{Suyu2010,KLiao2019}. However, the details of the observational uncertainty set up for the LSST lens are not yet known.
The current published data shows the inference of $D_{\Delta t}$ and $H_0$ approximation follow the Gaussian distribution~\cite{Birrer2019}. Therefore, in this paper, we only consider normal distribution for mock data. This assumption would not affect our main conclusion. More detailed discussions can be found in~\cite{KLiao2019,KLiao2019(2)}.

We now give the statistics for constraining $M_{B}$, $\alpha$ and $\beta$ in the two methods, respectively. In the ADD method, the statistical quantity can be expressed by using $\chi^{2}$:
\begin{equation}
\chi^{2}=\sum\limits_{i=1}^{N}\left[\frac{D^{A(i)}_{SN,l}-D^{A(i)}_{GL,l}}{\sigma_{D^{A}_{l}}^{i}} \right]^{2},
\end{equation}
where $D^{A(i)}_{SN,l}$ is the ith ADD term obtained from the SNe Ia data. $D^{A(i)}_{GL,l}$ is the ith data of ADD obtained from the strong lensing observations. N is the number of matched pairs that meet the screening criteria. The ith total error $\sigma_{D^{A}_{l}}^{i}$ can be written as
\begin{equation}
  \sigma_{D^{A}_{l}}^{i}=\sqrt{\sigma_{D^{A(i)}_{GL,l}}^{2}+\sigma_{D^{A(i)}_{SN,l}}^{2}},
\end{equation}
where $\sigma_{D^{A(i)}_{GL,l}}$ and $\sigma_{D^{A(i)}_{SN,l}}$ are ADD errors from gravitational
lensing and SNe Ia, respectively. Note that the latter depends on the parameters ($M_B, \alpha, \beta$).

In the TDD method, the statistical quantity can be written as
\begin{equation}
\chi^{2}=\sum\limits_{i=1}^{N}\left[\frac{D_{\Delta t, SN}^{(i)}-D^{(i)}_{\Delta t, GL}}{\sigma_{\Delta t}^{i}} \right]^{2},
\end{equation}
where $D_{\Delta t, SN}^{(i)}$ is the ith time delay distance term calculated from the supernova data according to Eq. 13 and $D_{\Delta t, GL}^{(i)}$ is the ith data of time delay distance obtained through observing gravitational lensing. N is the number of matched pairs that meet the screening criteria. The ith total error $\sigma_{\Delta t}^{i}$ can be written as
\begin{equation}
  \sigma_{\Delta t}^{i}=\sqrt{\sigma_{D_{\Delta t,GL}^{(i)}}^{2}+\sigma_{D_{\Delta t,SN}^{(i)}}^{2}},
\end{equation}
where $\sigma_{D_{\Delta t, GL}^{(i)}}$ and $\sigma_{D_{\Delta t, SN}^{(i)}}$ are the time delay distance errors obtained by gravitational lensing and the corresponding supernovae data, respectively.

We also consider the case where the universe is opaque. Such effect has been proposed to account for the dimming through dust distribution in the Milky Way, host galaxies, intervening galaxies and intergalactic media~\cite{Combes,Conley,Menard,Imara,McKinnon}. Cosmic opacity might result from some exotic mechanisms including the theory of gravitons~\cite{Chen}, Kaluza-Klein modes associated with extra-dimensions~\cite{Deffayet}, or a chameleon field~\cite{Khoury,Burrage}. The flux received by the observer will be reduced by the opacity depth factor, and the observed luminosity distance can be expressed by the opaque depth~\citep{Liao2015}:
\begin{equation}
  D^{L}_{SN,obs}=D^{L}_{SN,true}e^{\tau(z)/2},
\end{equation}
where $D^{L}_{SN,obs}$ is the observed luminosity distance from SN Ia, $D^{L}_{SN,true}$ is the true luminosity distance without the influence of opacity. In this paper, the opacity depth $\tau(z)$ is parameterized and can be written as
\begin{equation}
  \tau(z)=2\varepsilon z.
\end{equation}
Note that the distance information of the gravitational lensing is obtained by measuring the angular separation, regardless of the absolute intensity. That is, the distance measured by gravitational lensing are not biased even in the presence of opacity. Therefore, We also study the calibration by using ADD and TDD respectively under the influence of cosmic opacity.

\section{Simulations and results}
\label{Sec4}

To perform a study on the power of calibration, we take a flat $\Lambda$CDM universe with matter density $\Omega_M=0.3$ and Hubble constant $H_{0}=70 km s^{-1} Mpc^{-1}$ as our fiducial model in the following simulations.

Since this work aims at giving a prediction of constraints on $\alpha,\beta,M_{B},\varepsilon$ rather than using realistic data to get a result, we give an unbiased analysis reflecting an average constraining power by the following steps.
Firstly, on the basis of the distributions in JLA, we set the parameters ($\alpha,\beta,M_{B}$) and the theoretical observational quantities ($m_B, x, c$) of SNe Ia such that the luminosity distances of SNe Ia can be converted to the fiducial values. Secondly, we randomly select the redshifts of lenses and sources from Fig.1, then calculate the corresponding lensing distances, note that the number of matched pairs might be different for each selection. Thirdly, We perform the noise distribution by considering the uncertainty levels of supernovae data in JLA and $5\%$ uncertainties for lensing distances to generate the mock data. Fourthly, we do minimizations to find the best-fits of parameters ($\alpha, \beta, M_{B}$, $\varepsilon$) by using the minimization function in Python. Finally, we repeat the minimization process for 50,000 times under different noise realizations.

\begin{figure}
\includegraphics[width=8cm,angle=0]{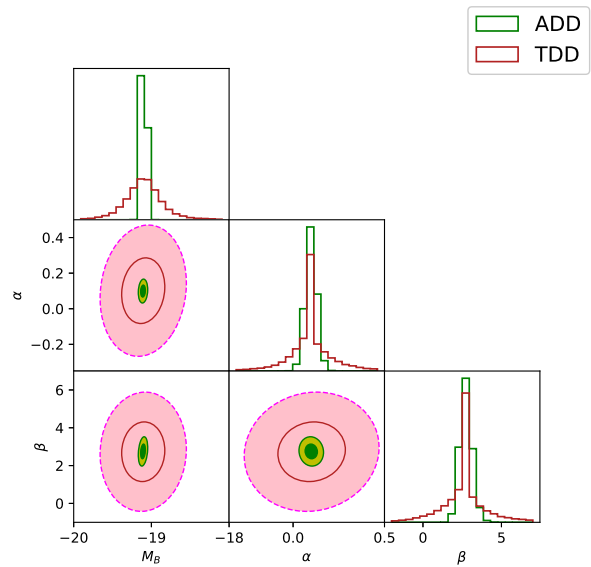}
\caption{In the case of no opacity, the 1-D and 2-D marginalized distributions and 1$\sigma$ and 2$\sigma$ constraint contours for SNe Ia nuisance parameters $(\alpha, \beta, M_{B})$, respectively. The green contours and red contours represent the constraint results for the ADD method and the TDD method, respectively.}
\label{result2}
\end{figure}

\begin{figure}
\includegraphics[width=8cm,angle=0]{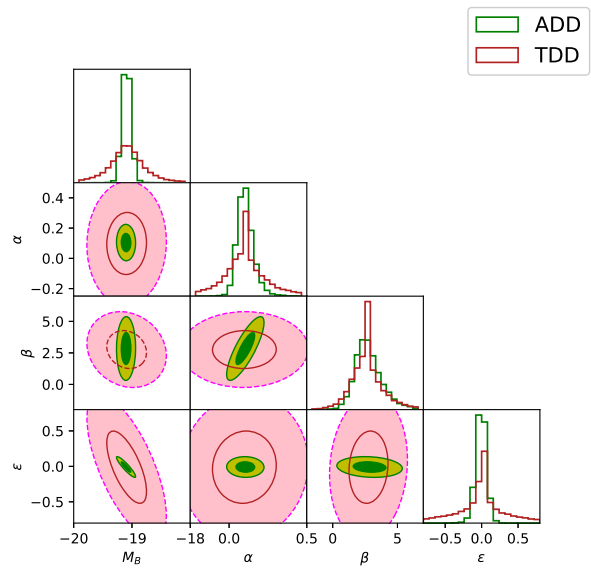}
\caption{The same as Fig.\ref{result2} but in the case of considering the cosmic opacity.}

\label{result3}
\end{figure}

We take all the best-fits from each minimization as the expected distributions of the parameters. For both methods, we show results that are not affected by opacity in Fig.\ref{result2}, and that considering the effect of opacity in Fig.\ref{result3}. The constraints of each parameter of supernovae are composed of one-dimensional distributions corner plots and two-dimensional constraint for the combination of two parameters, where the innermost contour and the outermost contour represent the $1\sigma$ and $2\sigma$ ranges, respectively.

The simulation results in the ADD method show that the $M_{B}$ uncertainty range of the supernova can be determined at 0.03 under $1 \sigma$ confidence level without the influence of opacity. Under the influence of opacity, the simulation results can determine the uncertainty range of $M_{B}$ at 0.08 $(1 \sigma)$. We also consider these two cases with the TDD method. The uncertainty range of $M_{B}$ is 0.26 $(1 \sigma)$ and 0.31 $(1 \sigma)$, respectively. For comparison,
we also consider the case when the lens data uncertainties are $10\%$. Results from the two methods are shown in table \ref{tab:final}.

\begin{figure}
\includegraphics[width=8cm,angle=0]{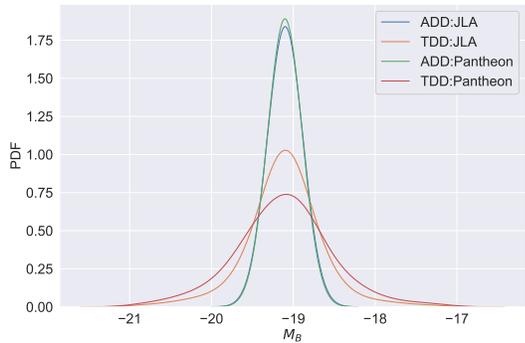}
\caption{The comparison of constraint results on $M_B$ between Pantheon sample and JLA sample with 5\% uncertainty of ADD and TDD.}

\label{result4}
\end{figure}

\begin{figure}
\includegraphics[width=8cm,angle=0]{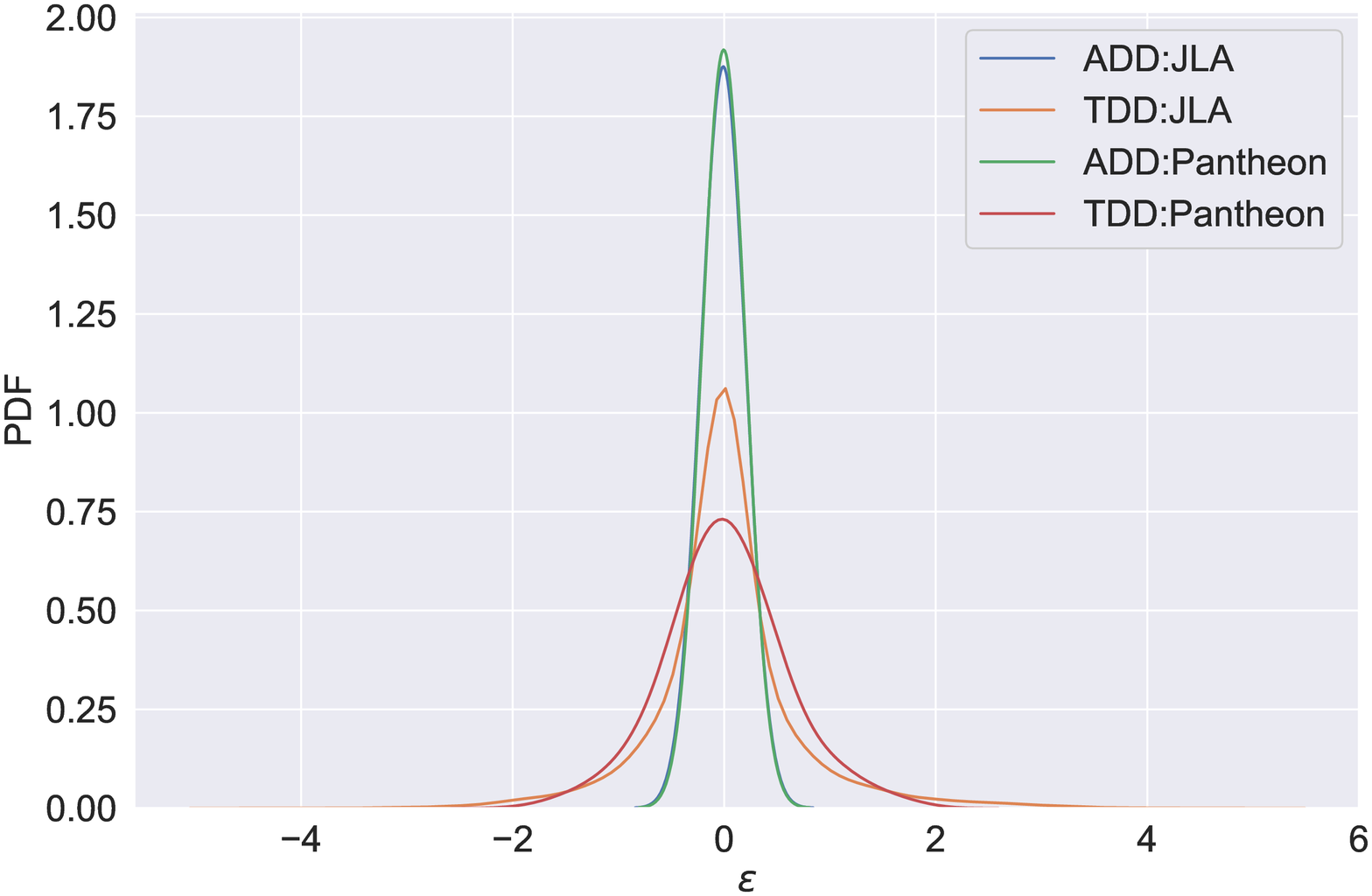}
\caption{The comparison of constraint results on $\epsilon$ between Pantheon sample and JLA sample with 5\% uncertainty of ADD and TDD.}

\label{result5}
\end{figure}

We further consider the Pantheon sample\cite{Scolnic2018} and compare the results with the JLA sample under the same conditions that the uncertainty of ADD and TDD are 5\%. For the Pantheon sample, there is no stretch-luminosity parameter $\alpha$ and color-luminosity parameter $\beta$, so we only consider absolute magnitude $M_{B}$ and cosmic opacity $\epsilon$.
As shown in Fig.\ref{result4} and Fig.\ref{result5}, both samples have almost the same results.

The results show that ADD method is much more powerful than TDD method. There are two reasons for this. First, and the most important reason, this is because the screening criteria we set for the TDD method will artificially weaken the constraints. There are $\sim53$ pairs of valid data that meet our screening criteria under the ADD method, but only $\sim2$ pairs for the TDD method due to the high-redshifts (typically 2-3) of the sources in LSST. Second, TDD method contains two distance errors from SNe Ia, while ADD method uses only one SN distance. Previous studies~\cite{KaiLiao2019} have shown that combining ADD and TDD can improve the ability to constrain the parameters. However in this work, the power of TDD is too weak. Jointing TDD and ADD does not affect, unless the power of TDD is at the same level as ADD. Therefore, we will not further study the effects of combining ADD and TDD.

Our study shows ADD method is excellent for calibrating SNe Ia. Nevertheless, we still incorporate the TDD method in this paper for completeness.

\begin{table}[ht]
\caption{}
\label{tab:final}
\centering
\begin{tabular}{ccccc}
\hline \hline
Methods & $M_{B}$  & $\alpha$ & $\beta$ & $\epsilon$ \\
\hline

ADD(5\%)  & $-19.10^{+0.03}_{-0.03}$ & $0.1^{+0.03}_{-0.03}$ & $2.7^{+0.40}_{-0.36}$ & \verb|\|\\
TDD(5\%)  & $-19.10^{+0.24}_{-0.24}$ & $0.1^{+0.11}_{-0.10}$ & $2.7^{+0.99}_{-1.13}$ & \verb|\|\\
ADD(10\%)  & $-19.10^{+0.04}_{-0.04}$ & $0.1^{+0.04}_{-0.04}$ & $2.7^{+0.54}_{-0.50}$ & \verb|\|\\
TDD(10\%)  & $-19.10^{+0.31}_{-0.31}$ & $0.1^{+0.16}_{-0.18}$ & $2.7^{+1.15}_{-1.66}$ & \verb|\|\\
\hline
ADD(5\%)  & $-19.10^{+0.06}_{-0.06}$ & $0.1^{+0.03}_{-0.03}$ & $2.7^{+0.37}_{-0.36}$ & $0.0^{+0.06}_{-0.06}$\\
TDD(5\%)  & $-19.10^{+0.31}_{-0.31}$ & $0.1^{+0.15}_{-0.15}$ & $2.7^{+1.20}_{-1.18}$ & $0.0^{+0.34}_{-0.39}$\\
ADD(10\%)  & $-19.10^{+0.11}_{-0.11}$ & $0.1^{+0.08}_{-0.07}$ & $2.7^{+1.89}_{-1.50}$ & $0.0^{+0.10}_{-0.10}$\\
TDD(10\%)  & $-19.10^{+0.39}_{-0.37}$ & $0.1^{+0.17}_{-0.17}$ & $2.7^{+1.38}_{-1.76}$ & $0.0^{+0.38}_{-0.42}$\\

\hline

\end{tabular}
\begin{flushleft}{Constraint results of supernova parameters for two methods with different degrees of uncertainty for the JLA sample. The top half of the table shows the results without opacity, and the bottom half considers results with opacity.}
\end{flushleft}
 \label{tab:data}
\end{table}

\section{Conclusion and Discussions}
\label{Sec5}

SNe Ia play an important role in modern astronomy, especially in measuring cosmological distances. However, some theories and observations introduced controversies about the absolute magnitude of supernovae and the distance relation from Cepheids.

In this paper, we propose a strategy to calibrate the absolute magnitude of supernovae by using two kinds of lensing distances. The simulation is based on the high-quality data available in the future LSST era. The results show that gravitational lensing systems can constrain the SN Ia parameters to a high precision. Compared with the TDD method, the ADD method is more powerful in constraining the parameters of supernovae. There are two main origins for the large uncertainty of the TDD method. On the one hand, it contains the error of two distances, which significantly increases the uncertainty. On the other hand, it is necessary to match the redshifts of both lens and source, resulting in a smaller amount of data that match successfully.
Applying TDD method for LSST lenses may not be the best idea in our matching method
since the source redshifts are usually too high compared with supernovae. Nevertheless, smoothing method like the Gaussian Process can make all the lensing systems whose source redshifts $<2$ available, which is worth trying in further studies.

For the absolute magnitude, Richardson et al. obtained the $M_{B}$ of $-19.25\pm0.2$ through a comparative study~\cite{Richardson}, which is much better than the $M_{B}=-19.16\pm0.76$ they obtained in 2001~\cite{Richardson2002}. The TDD method (5\%) can also obtain the same constraint results without considering the cosmic opacity. The results of the ADD method (5\%) are consistent with the results of the best-fit $\Lambda CDM$ parameters with the C11 sample $M_{B}=-19.16\pm0.03$ and slightly smaller than the JLA sample fitting results $M_{B}=-19.04\pm0.01$~\cite{Betoule}.
The best-fit parameter of $M_B$ found by combining BAO and SNe is $-19.12\pm0.03$~\cite{Macaulay}. The uncertainty of the $M_B$ which was constrained from the 110 Cepheid variables in the host galaxies of two recent SNe Ia (NGC 1309 and NGC 3021) is 0.05~\cite{Riess2009}. By using the new technique, Cepheid variables in 11 host galaxies of recent Sne Ia were observed in near-infrared, and 19 reliable SNe Ia samples were calibrated, resulting in the uncertainty of $M_B$ $\sim0.13$~\cite{Riess2016}.
The calibration uncertainty of $M_{B}$ of supernovae through gravitational wave events is $\sigma_{M_{B}}\simeq(0.1,0.2)$~\cite{zhao}. The uncertainty of $M_B$ calibrated from 1000 GW events is one order of magnitude smaller than the one calibrated with Cepheids~\cite{zhao}. Compared with the results of these methods, the gravitational lens can give more powerful constraints over a wider range of redshifts.

\section*{Acknowledgments}
We are grateful to the anonymous referee for helpful
comments. This work was supported by the National Natural Science Foundation of China (NSFC) No. 11973034 and 11603015.

\end{document}